\documentclass[letter]{ptptex}
\usepackage{wrapft}
\usepackage{color}

\usepackage{graphicx}

\pubinfo{Vol.~***, No.~*, May 2011} 

\title{
Effects of a New Triple-$\alpha$ Reaction on X-ray Bursts of a Helium Accreting Neutron Star
}

\author{
Yasuhide \textsc{Matsuo}$^{1,}$
\footnote{E-mail: matsuo@phys.kyushu-u.ac.jp}, 
Hideyuki \textsc{Tsujimoto}$^{1}$,
Tsuneo \textsc{Noda}$^{1}$,
Motoaki \textsc{Saruwatari}$^{1}$,
Masaomi \textsc{Ono}$^{1}$,
Masa-aki \textsc{Hashimoto}$^{1,}$\footnote{
E-mail: hashimoto@phys.kyushu-u.ac.jp}
and Masayuki Y. \textsc{Fujimoto}$^{2}$
}

\inst{
$^1$Department of Physics, Kyushu University, Fukuoka 812-8581, Japan
\\
$^2$Department of Physics, Hokkaido University,\\ 
Sapporo 060-0810, Japan
}

\abst{%
The effects of a new triple-$\alpha$ reaction rate (OKK rate) on the helium flash of a helium accreting neutron star in a binary system have been investigated.  Since the ignition points determine the properties of a thermonuclear flash of  type I
X-ray bursts, we examine the cases of different accretion rates, $dM/dt~(\dot{M})$, of helium
from $3\times10^{-10} M_{\odot}~\rm yr^{-1}$ to
$3\times10^{-8} M_{\odot}~\rm yr^{-1}$, 
which could cover the observed accretion rates. We find that  for the cases
of low accretion rates,  nuclear burnings are ignited at the
helium layers of rather low densities. As a consequence, helium deflagration would be triggered for all cases of lower accretion rate than $\dot{M}\simeq 3\times10^{-8} M_{\odot}~\rm yr^{-1}$.
We find that OKK rate 
is consistent with the available observations of the X-ray bursts on the helium accreting neutron star.
We advocate that the OKK rate is better than the previous rate for the astrophysical phenomena of
X-ray burst due to helium accretion.
}

\PTPindex{242, 421}   

\begin{document}
\maketitle

\noindent
1. {\it Introduction}

A new challenge has been given for the triple-$\alpha$ (3$\alpha$) reaction rate, which has been calculated by Ogata et al.\cite{rf:okk}
and found to be very large compared with the previous rate used so far.\cite{rf:ntm95,rf:angulo,rf:dotter,rf:morel} 
As a consequence, the new rate results in the helium 
($^4$He) ignition in the lower density/temperature
on the stellar evolution of low-, intermediate-, and high-mass stars\cite{rf:nh88,rf:hashi95}, accreting white dwarfs\cite{rf:nom82,rf:nom82b}, and accreting neutron stars.\cite{fuji81,miya85,peng2010}
Therefore, it is urgent to clarify quantitatively as possible as how the new rate affects the above astrophysical phenomena, because the rate plays the most fundamental role among the nuclear
burning in heavenly bodies and could do some role in the early universe, where
any terrestrial experiments for the 3$\alpha$ reaction are very difficult. In the present paper, we investigate the effects of  a newly calculated 3$\alpha$ reaction rate (OKK rate)\cite{rf:okk} on the helium flashes that occur at the bottom layers inside the accreting envelope of a neutron star. We can use 
the ignition curves  to find roughly when the nuclear ignition occurs. 
We note that X-ray bursts which have been mainly studied 
so far are limited to the combined burning of hydrogen and helium~\cite{lew93} and observational 
features such as light curves and burst energy have been qualitatively explained well.
Fujimoto et al.~\cite{Fuji84} have succeeded in simulating the X-ray bursts by solving the whole structure equations and clarified firstly the importance of the heat flow from the bottom of the accumulated layer.  
The application has been done for several X-ray burst observations of combined H and He burnings.
\cite{hana84,hanaf86,fuji87}
However,  there remained detailed comparison between observations and calculations related to
successful  bursts, very low accretion rates, and or superbursts (e.g., \citen{lew93,bild98}).

In the mean time, Fynbo et al.~\cite{rf:fynbo} (we call the reference as Fynbo) revised the 3$\alpha$ rate of NACRE~\cite{rf:angulo} based on new experiments
at high temperature of $T > 10^{9}$~K and with an artificial extrapolation to the
very low temperature region toward $T\sim10^7$~K. However, in the present case of 
accreting neutron stars we can regard the differences between NACRE and Fynbo
as unimportant , because the OKK rate is much larger for the temperature less than
$10^{8}$~K compared to the difference between NACRE and Fynbo.

It is noted that although nuclear burning strongly depends on the temperature, the density becomes very important at high densities of
$\rho \geq 10^6~\rm g~cm^{-3}$ and low temperatures of $T \leq 10^8$~K due to
 the screening effects.~\cite{rf:nom82}

 The new rate has been found to affect significantly the evolution of low-mass and intermediate stars\cite{rf:dotter,rf:morel}, 
where the evolutions from the zero-age  main sequence through the core He flash/burning 
from 1 $M_{\odot}$ to 10 $M_{\odot}$ have been investigated. 
From the HR diagram obtained with the OKK rate, the results
disagrees in detail with the observations.
  If the OKK rate in the temperature range of $10^8~{\rm K} < T < 2\times10^8~{\rm K}$ is exactly correct, we must invoke some new physical processes such as rotational mixing
and/or convection mechanism.

On the other hand, Saruwatari and Hashimoto~\cite{saru2010} have shown the
important difference between the ignition points calculated by the two rates for the helium accreting carbon-oxygen
white dwarf. They have concluded that for all the accretion rates, nuclear fuel  ignites
at the accumulated layers of helium  and a scenario of Type Ia supernovae changes
for low accretion rates of $\dot{M} < 4\times10^{-8}~M_{\odot}~\rm yr^{-1}$.
Related to this study, since the progenitor of Type Ia supernovae have been constructed with use of 
previous reaction rate, discussions concerning the origin of the supernovae that originate from the 
binary between a white dwarf and red giant  could be  changed.~\cite{rf:nom84}

Furthermore, simple helium ignition model on an accreting neutron star has been investigated.\cite{peng2010}
It is concluded that the OKK rate does not fit the observations of X-ray bursts due to the 
pure helium accretion but previous rate agrees with the observations. However, their model
is based on a model of one variable of column density
 and only steady state has been followed until the ignition.
It should be noted that the whole structure from the center to the surface of a neutron star must be solved to discuss the thermal evolution of accreting neutron stars. This has been stressed by Fujimoto et al.\cite{Fuji84}
 Therefore, it is desirable to include the
whole structure and follow the evolution of accreting neutron stars with physical processes
included~\cite{Fuji84}, if we try to elucidate the effects of
the OKK rate on the evolution and compare the results with the available observations. 
We note that for the combined burning of H and He, no difference between Fynbo and OKK
appears since the burst is triggered at higher temperature region of 
$T > 5\times10^8$~K.\cite{ww81}
In this letter, we present the results of the evolutionary calculation beyond the helium ignition of a helium accreting neutron star. We will show
the important role of both the 3-$\alpha$ reaction rates in the low temperature region and the properties of the crustal heating.~\cite{sato,cum06}

\noindent
2. {\it Evolution toward the explosive helium burning on the accreting neutron star}

Accreting neutron stars are considered to be  the origin of the type I X-ray bursts.\cite{Fuji84}  Accreting materials are  usually 
hydrogen and/or helium. Since the hydrogen is converted to  helium through steady hydrogen burning, helium is gradually accumulated on 
the neutron star
 and the deep layers become hot and dense. Helium flash triggered in the region composed of degenerate electrons could develop to  the 
dynamical stage, depending on the accretion rate $\dot{M}$.\cite{rf:tny84} The properties of ignition are determined from the thermal 
structure around  the
bottom of accreting layers. It has been found that unstable helium burning after the exhaustion of hydrogen results if the helium 
accretes in the range of
$2 \times 10^{-10} M_{\odot}~{\rm yr^{-1}} <\dot{\it M} < 10^{-9} M_{\odot}~{\rm yr^{-1}}$.~\cite{bild98}
 
We have computed the evolution of an accreting neutron star by the Henyey-type implicit-explicit method\cite{Fuji84} during the hydrostatic evolutionary stage of the accretion of helium, where the full set of general
relativistic equations of spherically symmetric stars is based on the formulation by Thorne.~\cite{thorne77} 
Physical inputs are the same as
Ref~\citen{Fuji84} except for the network of nuclear burning and additional heating from the crust.
We note that under the assumption of the spherically  symmetry, the basic four equations of neutron star evolution becomes the same type of those of the ordinary stellar evolution except for the effects of the general relativity. 

 Initial models are constructed by accreting helium on a neutron star of which the  gravitational mass and the radius  of the neutron star are $M=1.3~M_{\odot}$ and $R=8.1~\rm km$, respectively. 
We can obtain the initial models for each constant accretion rate shown in Table~\ref{Burst}.
after the steady state has been achieved 
as in Ref.~\citen{Fuji84}: the integrated energy flux of the non-homologous part for the gravitational energy release becomes negligible and therefore the thermal structure does not almost change. 
The masses have been gradually increased during the accretion toward the steady state
and the amounts are less than 1~\% of the total mass when the steady state has been
 achieved.

To follow the nuclear burning, the alpha-network has been implemented to obtain the nuclear energy
generation rate:~\cite{kurom}  the network consists
of $^4$He, $^{12}$C, $^{16}$O, $^{20}$Ne, $^{24}$Mg, $^{28}$Si, $^{32}$S, $^{40}$Ca, $^{44}$Ti, $^{48}$Cr,
 $^{52}$Fe, and $^{56}$Ni. We include and examine the two reaction rates of Fynbo and OKK for the 3$\alpha$ reaction,
 and the other reaction chain is selected from the faster reaction of either $(\alpha, \gamma)$ or $(\alpha, \rm p)$ followed by  the $(\rm p, \gamma)$ reactions. 
After the calculations of stellar structure have been converged, this network is operated 
to obtain the nuclear energy generation rate to calculate the next stellar structure as the same method in 
Ref.~\citen{hana84}.  The nuclear energy generation rate of the $3\alpha$ reaction is crucial to determine
the ignition condition and would be proportional to $T_9^{-3} \exp(-44.0/T_9)$~\cite{ww81} for the high temperature region of $T_9 > 0.1$. However, the temperature dependence cannot be written by a simple
analytical formula, that is,
\begin{equation}
\varepsilon _{3\alpha} \propto \rho^{2}\left(\frac{Y}{4}\right)^{3} f \langle 3\alpha \rangle(T), \label{epn3a}
\end{equation}
where $Y$ is the helium mass fraction and $f$ is the screening factor which is included by the same method in
Ref.~\citen{rf:ntm95}. As a function of $T$, $\langle 3\alpha \rangle(T)$  was obtained from the average with the Maxwell-Boltzmann distribution concerning the product of the cross section and the relative
 velocity.~\cite{rf:okk}

The importance of a crustal heating during the thermal evolution of neutron stars has been pointed
 out.~\cite{sato}  The heating rate is tabulated in Ref.~\citen{haen90} and the heating rate is
\begin{equation}
Q_i = 6.03\dot{M}_{-10}\frac{q_i}{1\rm{MeV}}~10^{33} ~\rm erg~s^{-1},
\end{equation}
where $\dot{M}_{-10}$ is accretion rate in units of $10^{-10}~M_{\odot}\rm{\,}yr^{-1}$  and
$q_i$ is deposited heat in MeV/nucleon whose representative value within the present
accretion rate in this paper would be evaluated to be 0.012~MeV/nucleon 
from the produced average heating rate: 
\begin{equation}
  <Q_i> = \int_{\rm{crust}} Q_i dm/\Delta~M, \qquad  \Delta M=\int_{\rm{crust}}dm,
\end{equation}
where $dm=4\pi r^2\rho dr$ and the integral covers the crust of the density range 
$10^{11}-10^{13}~\rm g~cm^{-3}$ and 
$\Delta~M \simeq 1.2\times10^{-4}~M_{\odot}$ with the thickness of 1~km.
We can easily include it just as the neutrino
energy loss rates and/or nuclear generation rates, that is, the rate of the crustal heating is added 
in addition to the nuclear energy generation rate, where we write the energy equation in Newtonian
approximation for simplicity:
\[
\frac{\partial L_r}{\partial M_r}=\varepsilon_n^{*} -\varepsilon_{\nu} -
T\left(\frac{\partial s}{\partial t}\right)_{M_r},
\]
where $L_r$ is the energy flow, $\varepsilon_n^{*}$ includes the nuclear energy generation rate,
crustal heating rate, $\varepsilon_{\nu}$ is the neutrino energy loss rate, $s$ is the specific
entropy.

\begin{figure}[t]
  \begin{minipage}{0.5\hsize}
    \begin{center}
      \includegraphics[scale=0.9]{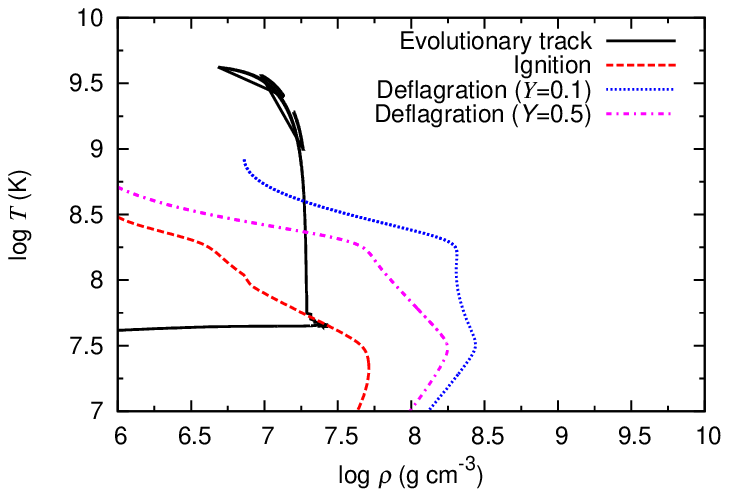}
    \end{center}
  \end{minipage}
  \begin{minipage}{0.5\hsize}
  \begin{center}
    \includegraphics[scale=0.9]{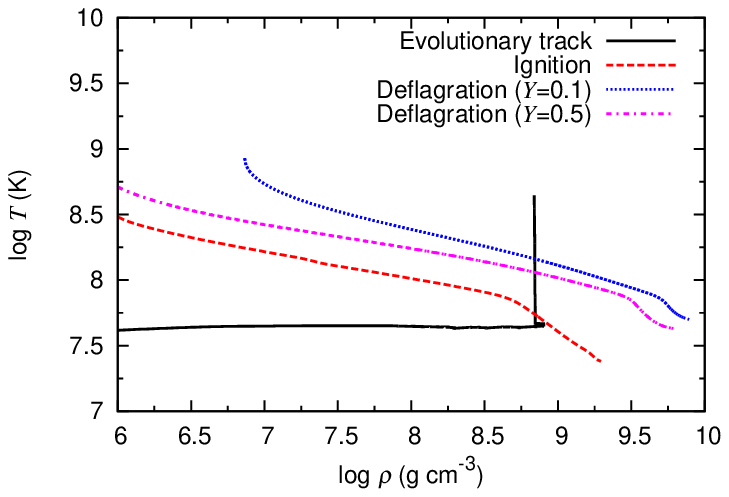}
  \end{center}
  \end{minipage}
  \caption{Evolutionary track of the temperature against the density at the bottom of the burning layer
           in  ${\dot M}{\;}={\;}3{\times}10^{-10} {\it{M_{\odot}}}{\,}{\rm{yr}}^{-1}$ for the $3\alpha$ reaction
           rate of OKK (left panel) and Fynbo (right panel). The ignition and deflagration curves are defined by 
           $\varepsilon_n = \varepsilon_{\rm rad}$ and Eq.(\ref{eq-deflag}), respectively.}
  \label{EvolTrack}
\end{figure}

We have selected six cases of accretion for two reaction rates (see Table~\ref{Burst}), where
the hight of the accreted layer is at most
some 10~m during the shell flash.
We have shown the evolutionary tracks in Fig.~\ref{EvolTrack} 
for the OKK rate (left panel) and the Fynbo rate (right panel) with the solid lines.
The tracks correspond to the bottom of the accretion layers, where the maximum
temperature results and shell flash should be triggered in the neighborhood  of the region.

Let us define the ignition and deflagration and/or detonation curves to show clearly the points of the helium ignition and the development of the shell flash. The density and the temperature in the below equations (4)-(8)
must correspond to the bottom of the burning layer and/or around that of the accreted layer.
The energy conservation law can be written as follows;
\begin{equation}
c_{P} \frac{dT}{dt} = {\varepsilon}_{n} - {\varepsilon}_{\rm{rad}}, \label{energyconserONEZONEEQ2}
\end{equation}
where $c_P$ is the specific heat at the
constant pressure and $\varepsilon_n$ is the nuclear generation rate of the  
3$\alpha$ reaction.
We approximate the radiative loss in the simple form,
\begin{equation}
{\varepsilon}_{\rm{rad}} =  \frac{4ac}{3{\kappa}} \frac{T^4}{{\sigma}^2}, \label{epsrnraddef}
\end{equation}
where $\kappa$ is the opacity, {\,}${\sigma}$ is the column density, $a$ is the radiation constant,
and $c$ is the speed of light.~\cite{hana}  
The column density can be obtained from the pressure
and the gravitational acceleration for a neutron star model (see Eq.~(\ref{ignitionp})).
The ignition curve is defined on the plane of (${\rho}$,$\it{T}$)
to satisfy the equality $\varepsilon_n = \varepsilon_{\rm rad}$. The ignition curves are drawn in
Fig.~\ref{EvolTrack} for the OKK rate and the Fynbo rate,
where we can infer that helium ignition should be triggered at the much lower density.

The dynamical timescale${\;}{\tau}_{\rm{dyn}}$ is defined as
\begin{equation}
{\tau}_{\rm{dyn}} = \frac{1}{{\;}\sqrt[]{\mathstrut 24{\pi}G{\rho}}},
\end{equation}
where $G$ is the gravitational constant.
On the other hand, the time scale ${\tau}_{n}$ of the increase of the temperature due to the nuclear burning is defined to be
\begin{equation}
{\tau}_{n} = \frac{c_{P}\it{T}}{{\varepsilon}_{n} }.
\end{equation}
The deflagration curve has been obtained from the below condition~\cite{rf:nom82}:
\begin{equation}
{\tau}_{n} {\;} = {\;} {\tau}_{\rm{dyn}}. 
\label{eq-deflag}
\end{equation}
If $\tau_n < \tau_{\rm dyn}$, deflagration wave should be originated,
 which could
develop and burn out the previously accumulated layers
(hereafter we call the wave the deflagration one).
The deflagration curves are also drawn on the plane of (${\rho}$,$\it{T}$) in Fig.~\ref{EvolTrack}.
We note that the curves depend significantly on the helium mass fraction designated by $Y$.

  Figure 1 shows the evolutionary track on the plane of  ($\rho, T$) toward the helium ignition
for the bottom of the burning layer  (around the bottom of the accreted layer)
and subsequent helium flash beyond the deflagration curves,
 where the deflagration curves of two cases
of helium mass fractions ($Y=0.1$ and $0.5$) are drawn with use of the OKK rate and Fynbo rate.
We can find that the helium ignition occurs at lower density regions by almost two  orders  of magnitude if the OKK rate is adopted.   
We note that for $\dot{M} < 3 \times 10^{-8}~M_{\odot}~\rm yr^{-1}$, the helium flash
develops to the dynamical stage. Therefore, our hydrostatic evolution code cannot follow the dynamical
stage which would lead to the deflagration and/or detonation.

\renewcommand{\arraystretch}{1.3}
\begin{table}[t]
  \begin{center}
    \caption{Energy releases $E_{\rm{burst}}$ per a burst for a fixed accretion rate 
             are given for two cases of OKK and Fynbo. The ignition pressure for each $\dot{M}$, $P_{\rm ign}$,
             corresponds to the maximum temperature layer inside the accretion layers whose location is around the
             bottom of the accumulated ones.}
    \label{Burst} 	
    \begin{tabular}{c|c|c|c} \hline \hline
reaction rate & \lower.1ex\hbox{$\dot{M}$}  [$M_{\odot}~{\rm yr^{-1}}$] & $ \log P_{\rm ign}[\rm{dyn}~cm^{-2}]$ & $E_{\rm{burst}}[\rm{erg}]$   \\ \hline
OKK    &  $3{\times}10^{-10}$  &  24.34 & $4.98{\times}10^{40}$ \\ 
       &  $8{\times}10^{-10}$  &  23.07 & $2.72{\times}10^{39}$ \\ 
       &  $1{\times}10^{- 9}$  &  23.04 & $2.52{\times}10^{39}$ \\ 
       &  $5{\times}10^{- 9}$  &  23.03 & $2.47{\times}10^{39}$ \\ 
       &  $1{\times}10^{- 8}$  &  22.90 & $1.80{\times}10^{39}$ \\
       &  $3{\times}10^{- 8}$  &  22.68 & $1.09{\times}10^{39}$ \\ \hline
Fynbo  &  $3{\times}10^{-10}$  &  26.47 & $6.84{\times}10^{42}$ \\ 
       &  $8{\times}10^{-10}$  &  24.35 & $5.13{\times}10^{40}$ \\ 
       &  $1{\times}10^{- 9}$  &  24.31 & $4.73{\times}10^{40}$ \\ 
       &  $5{\times}10^{- 9}$  &  23.83 & $1.51{\times}10^{40}$ \\ 
       &  $1{\times}10^{- 8}$  &  23.34 & $5.07{\times}10^{39}$ \\ 
       &  $3{\times}10^{- 8}$  &  22.68 & $1.09{\times}10^{39}$ \\ \hline \hline
   \end{tabular}
  \end{center}
\end{table}

\bigskip
\noindent
3. {\it Comparison with the observations}

It would be useful to judge which $3\alpha$ rate is more consistent with available observations. 
In general, it is very difficult to follow in the hydrodynamical calculations
the helium flash until the fuel of helium depletes as described in the
previous section.
However, using the results of evolutionary calculations obtained in the 
previous section, we can compare qualitatively the theoretical calculations with the observations.
We choose the layer of the helium ignition from which the pressure, the density, 
the temperature, and the composition of helium are obtained. 
We define the density ${\rho}_{\rm{ign}}$ and the pressure $P_{\rm{ign}}$ for the layer when the helium ignition occurs.
We can get ${\rho}_{\rm{ign}}$ from the crossing point between the evolutionary track of the bottom of the accreting layer and the
ignition curve. As a consequence, $P_{\rm{ign}}$ is obtained from the equation of state.
Since the accumulated layers are well approximated by the plane parallel configuration,
we can adopt the approximation to estimate the accumulated mass with a enough approximation;
It is written in the spirit of the plane parallel  model as follows:~\cite{hana82,hanaf82,hashi}
\begin{equation}
P_{\rm{ign}} {\,}={\,} g_{s} {\sigma}_{\rm{ign}} \label{ignitionp},
\end{equation}
where $g_s$ is the gravitational acceleration and
${\;}{\sigma}_{\rm{ign}}$ is the the column density at the ignition point.
The energy release $E_{\rm{burst}}$ can be obtained under the assumption that the accumulated layers
on the ignition layer are completely burnt out:
\vspace{-6pt}
\begin{eqnarray}
E_{\rm{burst}} &{\,}={\,}& {\Delta}M_{\rm{ign}} {\;} Q_{\rm{nuc}} {\;}/ ({1 + z_{s}}),  \\
		    &{\,}={\,}& 4{\pi}R^2{\sigma}_{\rm{ign}} {\;} Q_{\rm{nuc}} {\;}/ ({1 + z_{s}}). 
\label{burste} 
\end{eqnarray}
Here, ${\Delta}M_{\rm{ign}}$ is the total rest mass of the accreted layers,
the gravitational redshift, $z_{s}{\!}={\!}{(1-2GM_{t}/c^{2}R)^{-1/2}}~{-1}{\;}{\simeq}{\;}0.38$, and $Q_{\rm{nuc}}$
is the nuclear energy release per nucleon ($Q_{\rm{nuc}} = 1.6 {\;} \rm{MeV/nucleon}$),
where all fuel of helium are assumed to be burnt into iron.
For our model of the neutron star,  since $g_s = GM(1+z_s)/R^2$, we get $\log g_s = 14.56$.

\begin{figure}[t]
  \begin{minipage}{1.0\hsize}
  \begin{center}
    \includegraphics[scale=0.6]{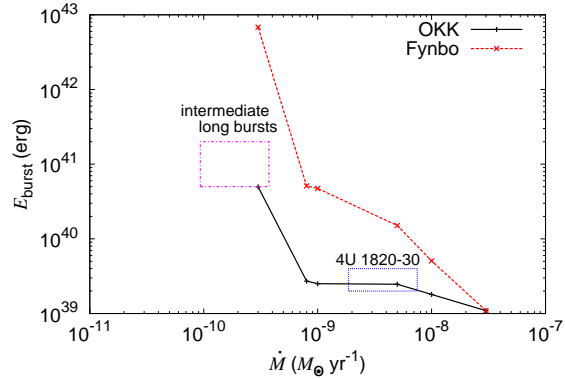}
  \end{center}
  \end{minipage}
  \caption{Energy release of $E_{\rm{burst}}$ against the accretion rate ${\dot M}$ for two cases of OKK and Fynbo. 
Observational values from different astrophysical objects which have been identified to cause type I X-ray
bursts due to the pure helium accretion, are plotted as was done in Ref.~\citen{peng2010}.}
  \label{EburstACCrate}
\end{figure}

In Table~\ref{Burst}, we show $P_{\rm{ign}}$ and $E_{\rm{burst}}$ in each accretion rate for the cases of
OKK and Fynbo.
In Fig.~\ref{EburstACCrate}, we show $E_{\rm{burst}}$ as a function of accretion rates
${\dot M}$, where are shown the two observations of
type I X-ray bursts, 4U1820-30 (labeled by 4U1820-30), SLX 1737-282, SLX 1735-269, and 2S 0918-549 (labeled
by intermediate long bursts).
Their accreted matter from companions has been known to be pure helium
\cite{1987ApJ...314..266H, 2002A&A...389L..43I, 2008A&A...484...43F, 2005A&A...434.1069M, 2005A&A...441..675I, 2011A&A...525A.111I}.
As is written in  Ref.~\citen{peng2010}, the uncertainties of accretion rates and burst energies are indicated by boxes.
For these three observations, we can see
from Fig.~\ref{EburstACCrate}, results by OKK are consistent with the observations
if we choose the accretion rate of $\dot{M} \leq  3\times10^{-10} M_{\odot}~\rm yr^{-1}$. 

\bigskip

\noindent
4. {\it Discussion and conclusions}

The ignition densities that determine the triggering mechanism on helium accreting neutron stars will be changed significantly 
if we adopt the new 3$\alpha$ reaction rate for the low accretion rate of $\dot{M}< 10^{-8} M_{\odot}~\rm yr^{-1}$.  
Until the OKK rate has been presented, the method by Nomoto\cite{rf:nom82} has been used of which
a simple extrapolation of the Breit-Wigner type function to the low-temperature side is adopted,
where it has been advocated that the non-resonant 3$\alpha$ reaction  is crucial in determining a helium ignition of compact stars 
for the low accretion rate. 
However, the microscopic calculation for the three
body problem has been found to be crucial in evaluating the 3$\alpha$ reaction rate. 
In the present paper, we have examined the effects of the OKK rate on the very low temperature
site of astrophysics.
Although our results seem to contradict with those of Peng and Ott~\cite{peng2010}, 
we may ascribe it to the following. 
First, we have adopted the rate of the crustal heating from Ref.~\citen{haen90}.
Second, they have used the mass of $1.4~M_{\odot}$ and the radius of 10~km for a neutron star.
Third, while we have followed the evolution of an accreting neutron star from the center to
the surface of the star, they have solved only steady-state
equations above the crust as a function of a single dependent variable of the column density. 
As a consequence, it has been written that the outward
energy release due to the crustal heating was taken to be a free parameter as a boundary condition,
which would be inappropriate in the calculations of accreting neutron stars.

\begin{figure}[tbp]
  \begin{center}
    \includegraphics[scale=0.8]{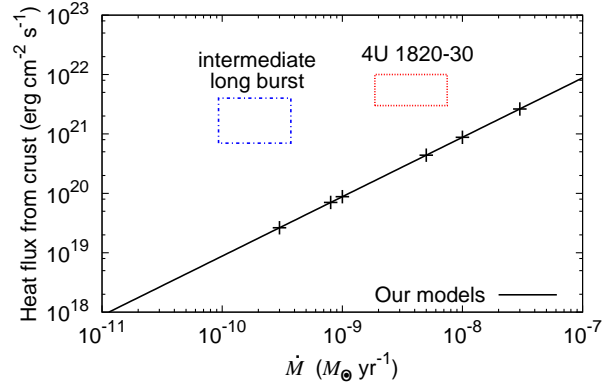}
  \end{center}
  \caption{Energy flux in units of $\rm erg~cm^{-2}s^{-1}$ from the crust against accretion rates with
           use of the table \cite{haen90} (solid line) and the two boxes are taken from Ref.~\citen{peng2010}.}
  \label{cheat}
\end{figure}

The most important difference comes from the treatment of the crustal heating.
In the present case, the energy flux ($\rm erg~cm^{-2}s^{-1}$) from the crust is 
shown in Fig.~\ref{cheat}. The average
energy per unit mass from the crust can be evaluated to be $\sim$~0.01~MeV/nucleon as shown
in the previous section. 
On the contrary, the rate they have adopted could be nearly 0.1~MeV/nucleon~\cite{cum06}.
These differences reflect the calculations of the heat flux as shown in
Fig.~\ref{cheat}, where the solid line represents heat flux in our calculations and
the two boxes are taken from Ref.~\citen{peng2010}.
It is clear that the heat flux from the crust in the present study is 1-2 orders of magnitude lower than the case in Ref~\citen{peng2010}.
As a consequence,
our initial models just before the accretion starts might have lower temperatures compared to those
used in Ref.~\citen{peng2010}. The most significant difference would come from the initial model and therefore,
the location of the ignition point on the $(\rho, T)$ plane. While the bottom of the accreted layers in our
 initial model for $\dot{M}=3\times10^{-10}M_{\odot}~\rm yr^{-1}$ has the temperature of 
 $T=1.6\times10^7$~K (see Fig.~\ref{EvolTrack}.),
we can infer for the case of Ref.~\citen{peng2010} the temperature could be rather high.
From the column density obtained from the box of the observation of intermediate long
bursts (see Fig.~2 in their paper), we can get the ignition temperature of
 $T=(1-2)\times10^8$~K for the Fynbo rate. If the initial temperature of their calculation take nearly the
same temperature, which is the common feature of the accreting compact objects before the 
beginning of the shell flashes\cite{sn80}, ignition density is found to be in the neighborhood  of that
 for the OKK rate.   Contrary to our calculations which are the results
of stellar evolution and include all the necessary physical processes, their calculations depend essentially
on the numerically convenient boundary condition at the assumed surface of the crust.

We have shown that OKK rate must change the scenario of Type Ia supernovae and type I X-ray bursts
for the low accretion rates, where the temperature at the ignition is less than $10^8$~K. 
 Since the rate was calculated under the situation of three $\alpha$ particles in vacuum, application of
the  OKK rate is doubtful against the high density. As seen from Fig. 3 in Ref.~\citen{rf:okk}, 
the discretized continuum wave functions of the two $\alpha$ particle system extend to $5\times10^3$~fm. 
Therefore, in the medium of stellar plasma, effects more than three $\alpha$ particles cannot be negligible for $\rho > 10^6~\rm g~cm^{-3}$.\cite{ogata}  For example,
the reaction rate is to be calculated with the inclusion of an appropriated screening potential.
If the effects of many body interactions are included, the OKK rate could be decreased by orders of magnitude.
We may constrain the 3$\alpha$ reaction rate from the astrophysical observations of type I X-ray
bursts. 

As example of constraint the OKK rate, if we adopt the upper values of the observational box of
 $E_{\rm burst}$ in  Fig.~\ref{EburstACCrate} for
$\dot{M}=3\times10^{-10} M_{\odot}~\rm yr^{-1}$, we obtain $\log P_{\rm ign}\sim24.9$ 
from (\ref{ignitionp})--(\ref{burste}).
Since the evolutionary paths are almost the same for the two $3\alpha$ rates until the ignition points (see
Figs.~\ref{EvolTrack}),
the corresponding ignition density with the evolutionary paths of OKK
becomes in the range of $\log\rho_{\rm ign} = 7.7-7.8$ for the OKK rate reduced by a factor of $10^{2-3}$.  
We note that the properties of shell flash depend to some extent on the structure
of a neutron star (mass and radius); Since $E_{\rm burst}$ is proportional to
$R^4/M$, a rather hard EOS ($R=12$~km and $M=2.0~M_{\odot}$) gives three times
larger value compared to our case, which would be preferable for the OKK rate. 
On the other hand, it depends on the non-standard cooling processes:
if we use the non-standard cooling processes such as a pion-condensation~\cite{Fuji84}, 
our estimates would be changed significantly.

In conclusion, we have found that
the new rate affects the helium ignition in accreting neutron stars, where 
for lower accretion rates, helium burns at lower densities and 
 temperatures, which should change the epoch of the formation of a helium deflagration wave and the modeling of type I X-ray bursts.\cite{rf:ntm95}
In particular, we consider that the crustal heating affects the ignition properties
significantly. Furthermore,
the mechanism behind superbursts should be studied again using the new
rate, because the amount of $^{12}$C could play a crucial role in inducing the
superbursts.\cite{rf:bild01}

\section*{Acknowledgements}
This work has been supported in part by a Grant-in-Aid for Scientific Research (19104006, 21540272) of the Ministry of Education, Culture, Sports, Science and Technology of Japan.


\begin{thebibliography}{99}

\bibitem{lew93}
       W H. G. Lewin, J. van Paradjis and R. E. Taam, Space Sci. rev. {\bf 62} (1993), 223.
\bibitem{hanaf86}
      T. Hanawa and M. Y. Fujimoto,
       Publ. Astron. Soc. Japan {\bf 38} (1986), 13.
\bibitem{fuji87} 
      M. Fujimoto, M. Sztajno, W.H. Lewin, J. v. Parajijs, \AJ{319,1987,902}.
\AJ{253,1982,798}.
\bibitem{rf:okk} 
		K. Ogata, M. Kan and M. Kamimura, \PTP{122,2009,1055}.\\
		K. Kan et al., JHP-Supplement-20 (1996), 204. \\ 
		K. Kan, Master thesis 1995 in Kyushu University, unpublished.
\bibitem{rf:ntm95} K. Nomoto, F.-K. Thielemann and S. Miyaji, \JL{Astron. Astrophys.,149,1985,239}.
\bibitem{rf:angulo} C. Angulo et al., \JL{Nucl. Phys. A,656,1999,3}.
\bibitem{rf:dotter} A. Dotter and B. Paxton, \JL{Astron. Astrophys.,507,2009,1617}.
\bibitem{rf:morel} P. Morel, J. Provost, B. Pichon, Y. Lebreton and F. Th${\rm {\acute e}}$venin, \JL{Astron. Astrophys.,520,2010,A41}.
\bibitem{rf:nh88} K. Nomoto and M. Hashimoto, Phys. Rep. \textbf{163} (1988), 13.
\bibitem{rf:hashi95} M. Hashimoto, Prog. Theor. Phys. \textbf{94} (1995), 663. 
\bibitem{rf:nom82} K. Nomoto, \AJ{253,1982,798}.
\bibitem{rf:nom82b}  K. Nomoto, \AJ{257,1982,780}.
\bibitem{fuji81} M. Y. Fujimoto, T. Hanawa and S. Miyaji, \AJ{247,1981,267}.

\bibitem{miya85} S. Miyaji and K. Nomoto, \JL{Astron. Astrophys.,152,1985,33}.
\bibitem{peng2010} F. Peng and C. D. Ott, \AJ{725,2010,309}.
\bibitem{rf:fynbo} H. O. U. Fynbo et al., \JL{Nature,433,2005,136}. 
\bibitem{saru2010} M. Saruwatari and M. Hashimoto, \PTP{124,2010,925}.
\bibitem{rf:nom84} K. Nomoto, F.-K. Thielemann and K. Yokoi, \AJ{286,1984,644}.
\bibitem{Fuji84} M. Y. Fujimoto, T. Hanawa, I. Iben Jr. and M. B. Richardson, \AJ{278,1984,813}.
\bibitem{sato} K. Sato, \PTP{62,1979,957}.
\bibitem{cum06} A. Cumming, J. Macbeth, J. J. M. in't Zand and D. Page, \AJ{646,2006,429}.
\bibitem{rf:tny84} F.-K. Thielemann, K. Nomoto and K. Yokoi, \JL{Astron. Astrophys.,286,1984,644}.
\bibitem{bild98}  L. Bildsten, In NATO ASIC Proc. 515: The Many Faces of Neutron Stars, ed. R. Buccheri,
           J. van Parajis, M. A. Alpar (Dordrecht : Kluwer), (1998), 419.
           
\bibitem{thorne77} K. S. Thorne, \AJ{212,1977,825}.
\bibitem{kurom} 
      R. Kuromizu, O. Koike, M. Hashimoto and K. Arai,
      Physics Reports of Kumamoto University {\bf 11} (2002), 197.
\bibitem{hana84}
      T. Hanawa and M. Y. Fujimoto,
       Publ. Astron. Soc. Japan {\bf 36} (1984), 199.
\bibitem{ww81}
      R. K. Wallance and S. E. Woosley, \AJ{45,1981,389}.
\bibitem{haen90} P. Haensel and J. L. {Zdunik}, \JL{Astron. Astrophys.,227,1990,431}.
\bibitem{hana}
      T. Hanawa, D. Sugimoto and M. Hashimoto,
       Publ. Astron. Soc. Japan {\bf 35} (1983), 491.
\bibitem{hana82}
      T. Hanawa and D. Sugimoto,
       Publ. Astron. Soc. Japan {\bf 34} (1982), 1.
\bibitem{hanaf82}
      T. Hanawa and M. Fujimoto,
       Publ. Astron. Soc. Japan {\bf 34} (1982), 495.
\bibitem{hashi}
      M. Hashimoto, T. Hanawa and D. Sugimoto,
      Publ. Astron. Soc. Japan {\bf 35} (1983), 1.
\bibitem{1987ApJ...314..266H} F. Haberl, L. Stella, N. E. White, W. C. Priedhorsky and M. Gottwald, \AJ{314,1987,266}.
\bibitem{2002A&A...389L..43I} J. J. M. in't Zand et al., \JL{Astron. Astrophys.,389,2002,L43}.
\bibitem{2008A&A...484...43F} M. Falanga, J. Chenevez, A. Cumming, E. Kuulkers, G. Trap and A. Goldwurm, \JL{Astron. Astrophys.,484,2008,43}.

\bibitem{2005A&A...434.1069M} S. Molkov, M. Revnivtsev, A. Lutovinov and R. Sunyaev, \JL{Astron. Astrophys.,434,2005,1069}.
\bibitem{2005A&A...441..675I} J. J. M. in't Zand, A. Cumming, M. V. van der Sluys, F. Verbunt and O. R. Pols, \JL{Astron. Astrophys.,441,2005,675}.
\bibitem{2011A&A...525A.111I} J. J. M. in't Zand, D. K. Galloway and D. Ballantyne, \JL{Astron. Astrophys.,525,2011,A111}.
\bibitem{sn80} D. Sugimoto and K. Nomoto, Space Sci. Rev. {\bf 25} (1980), 155.
\bibitem{ogata} K. Ogata, private communication.
\bibitem{rf:bild01} A. Cumming and L. Bildsten, \AJ{559,2001,L127}.



 
\end{thebibliography}
\end{document}